\documentclass[aps,prl,reprint,groupedaddress]{revtex4-1}

\newcommand{\be}{\begin{equation}}
\newcommand{\ee}{\end{equation}}

\usepackage{verbatim}
\usepackage{latexsym,amssymb,float}
\usepackage{setspace}
\usepackage{graphicx}
\usepackage{epsfig}
\usepackage{amsmath}
\usepackage{bm}

\begin{document}

\title{Orbital superconductivity, defects and pinned nematic fluctuations in the doped iron chalcogenide FeSe$_{0.45}$Te$_{0.55}$}
\author{Saheli Sarkar$^1$}
\author{John Van Dyke$^{1,2}$}
\author{Peter Sprau$^{3,4}$}
\author{Freek Massee$^{3,4,5}$}
\author{Ulrich Welp$^6$}
\author{Wai-Kwong Kwok$^6$}
\author{J.C. Seamus Davis$^{3,4,7,8}$}
\author{Dirk K. Morr$^1$}

\affiliation{$^1$ University of Illinois at Chicago, Chicago, IL 60607, USA}

\affiliation{$^2$ Department of Physics and Astronomy, Iowa State University, Ames, Iowa 50011, USA}

\affiliation{$^3$ LASSP, Department of Physics, Cornell University, Ithaca, NY 14853, USA}

\affiliation{$^4$ CMPMS Department, Brookhaven National Laboratory, Upton, NY 11973, USA}

\affiliation{$^5$ Laboratoire de Physique des Solides (CNRS UMR 8502), B\^{a}timent 510, Universit\'{e} Paris-Sud/Universit\'{e} Paris-Saclay, 91405 Orsay, France}

\affiliation{$^6$  Materials Science Division Argonne National Laboratory, Argonne, IL 60439, USA}

\affiliation{$^7$ School of Physics and Astronomy, University of St. Andrews, Fife KY16 9SS, Scotland, UK }

\affiliation{$^8$ Tyndall National Institute, University College Cork, Cork T12R5C, Ireland}

\date{\today}

\begin{abstract}
We demonstrate that the differential conductance, $dI/dV$, measured via spectroscopic imaging scanning tunneling microscopy in the doped iron chalcogenide FeSe$_{0.45}$Te$_{0.55}$, possesses a series of characteristic features that allow one to extract the orbital structure of the superconducting gaps. This yields nearly isotropic superconducting gaps on the two hole-like Fermi surfaces, and a strongly anisotropic gap on the electron-like Fermi surface. Moreover, we show that the pinning of nematic fluctuations by defects can give rise to a dumbbell-like spatial structure of the induced impurity bound states, and explains the related $C_2$-symmetry in the Fourier transformed differential conductance.

\end{abstract}


\maketitle

Identifying the electronic structure of the iron-based superconductors has remained one of the most important open challenges in determining the underlying superconducting pairing mechanism. There exists strong evidence for a $s_\pm$-wave symmetry of the superconducting order parameter \cite{Ste11,Hir11}, arising from the pairing between electron-like and hole-like Fermi surface pockets that is mediated by magnetic fluctuations \cite{Maz08,Chu08,Gra09,Si16}. However, the existence of nematicity in the iron chalcogenide superconductor FeSe \cite{Fer14}, and the observation of high temperature superconductivity in iron chalcogenides without hole pockets \cite{Qian11,Niu15,Zhao16}, has recently cast doubt on such a simple picture, giving rise to the proposal of orbital selective superconducting pairing \cite{Sprau16a,Sprau16b,Yin11,Med14}. Moreover, recent scanning tunneling spectroscopy experiments \cite{Sin15} have reported that even outside the nematic phase in doped FeSe$_{0.4}$Te$_{0.6}$, the electronic $C_4$-symmetry is broken, leading to an approximate $C_2$-symmetry in the Fourier transformed differential conductance. Whether such a symmetry breaking arises from the pinning of dynamic nematic fluctuations, or from orbital selective Mottness, is presently unclear.

In this Letter, we provide insight into these open questions by analysing the results of spectroscopic imaging scanning tunneling microscopy (STM) experiments in the doped iron chalcogenide FeSe$_{0.45}$Te$_{0.55}$. In particular, we demonstrate that the differential conductance, $dI/dV$, possesses a series of characteristic features that allows us to extract not only the orbital structure of the superconducting order parameter, but also its momentum dependence along the three Fermi surface sheets, yielding two nearly isotropic superconducting gaps for the hole-like Fermi surfaces, and a highly anisotropic gap for the electron-like Fermi surface. Moreover, we show that the pinning of nematic fluctuations gives rise to the experimentally observed dumbbell-like spatial structure of the induced impurity states, and leads to the nearly $C_2$-symmetric Fourier transformed differential conductance observed experimentally \cite{Sin15}.

Starting point for the analysis of the differential conductance, $dI/dV$, measured in the superconducting state of FeSe$_{0.45}$Te$_{0.55}$, is a five-orbital, tight-binding Hamiltonian $H=H_0+H_{SC}$ \cite{Gra09} which in real space is given by
\begin{align}
 H_{0} =& -\sum_{{\bf r},{\bf r'},\sigma}\sum_{\alpha,\beta=1}^{5} t^{\alpha\beta}_{\bf r,r'} c_{{\bf r},\alpha,\sigma}^{\dagger} c_{{\bf r'},\beta,\sigma} \nonumber \\
 H_{SC} & = - \sum_{\langle {\bf r},{\bf r'} \rangle }\sum_{\alpha=1}^{5}
 I_{{\bf r,r'}}^{\alpha} \ c_{{\bf r},\alpha,\uparrow}^{\dagger}c_{{\bf r'},\alpha,\downarrow}^{\dagger}c_{{\bf r},\alpha,\downarrow}c_{{\bf r'},\alpha,\uparrow} \ .
 \label{eq:Hamiltonian}
\end{align}
Here $\alpha,\beta=1,...,5$ are the orbital indices corresponding to the $d_{xz}$-, $d_{yz}$-, $d_{x^2-y^2}$-, $d_{xy}$-, and $d_{3z^2-r^2}$-orbitals, respectively,  $-t^{\alpha\beta}_{\bf r,r'} $ represents the electronic hopping amplitude between orbital $\alpha$ at site ${\bf r}$ and orbital $\beta$ at site ${\bf r'}$, and $c_{{\bf r},\alpha,\sigma}^{\dagger} (c_{{\bf r},\alpha,\sigma})$ creates (annihilates) an electron with spin $\sigma$ at site ${\bf r}$ in orbital $\alpha$. To obtain a superconducting order parameter with $s_{\pm}$-wave symmetry \cite{Maz08,Hir11}, we take $H_{SC}$ to describe superconducting intra-orbital pairing between next-nearest neighbor Fe sites (in the 1 Fe unit cell), with $I_{{\bf r,r'}}^{\alpha}$ being the pairing interaction. Using a mean-field decoupling of $H_{SC}$, we obtain
\begin{align}
H_{SC}^{MF} & = \sum_{\langle {\bf r},{\bf r'} \rangle }\sum_{\alpha=1}^{5} \Delta_{\alpha \alpha} c_{{\bf r},\alpha,\uparrow}^{\dagger} c_{{\bf r'},\alpha,\downarrow}^{\dagger} + H.c.
\end{align}
We extract the hopping amplitudes by fitting the momentum resolved bandstructure in the normal state of FeSe$_{0.42}$Te$_{0.58}$ obtained in angle-resolved photoemission spectroscopy (ARPES) experiments \cite{Tam10}, as shown in Fig.~\ref{fig:ARPES}(a) [their values are given in Tables I and II of the supplemental material (SM) Sec.I].
\begin{figure}[!ht]
\begin{center}
\includegraphics[width=8cm]{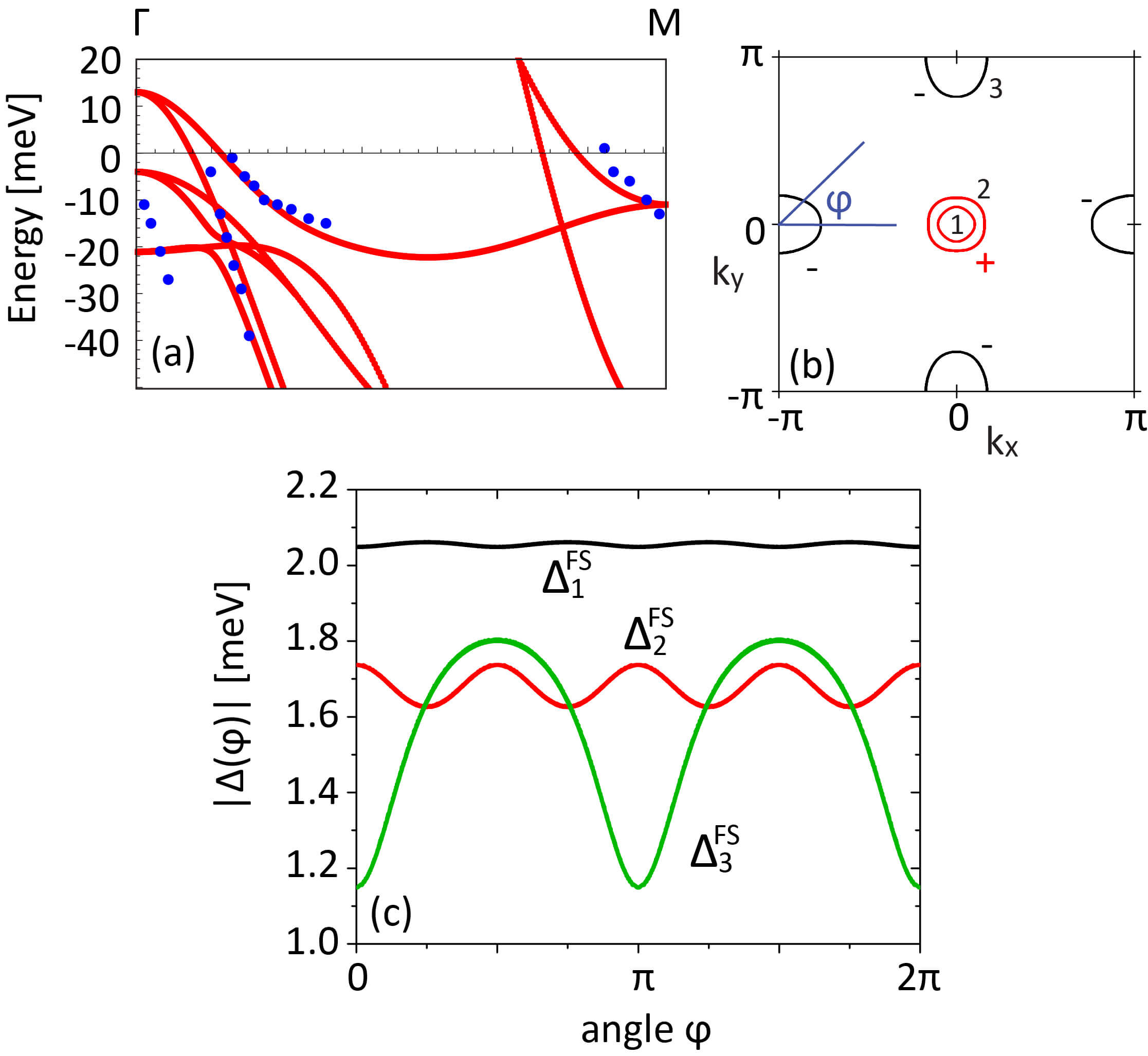}
\caption{(a) Theoretical fit of the experimentally determined energy dispersion of FeSe$_{0.42}$Te$_{0.58}$ \cite{Tam10} in the 2-Fe/cell Brillouin zone (BZ).  The blue dots represent the experimentally determined dispersion \cite{Tam10}, and the red lines represent the theoretical fit. (b) Plot of the three Fermi surface sheets resulting from the fit presented in (a) with two hole-like FS pockets around $\Gamma$ [$(0,0)$] point and one electron-like FS pocket around $M$ [$(\pm \pi,0), (0,\pm \pi)$] points in the 1Fe BZ. "$\pm$" indicate the phase of the superconducting $s_\pm$-wave order parameter. (c) Superconducting gap along the three Fermi surfaces as a function of angle $\varphi$ measured with respect to the $x$-axis.
}
\label{fig:ARPES}
\end{center}
\end{figure}
FeSe$_{0.42}$Te$_{0.58}$ possesses three Fermi surface (FS) sheets, with two hole-like FS pockets closed around the $\Gamma$ point (FS 1 and 2) and one electron-like FS pocket around the $M$ points [$(\pm \pi,0), (0,\pm \pi)$] (FS 3) in the 1Fe/cell Brillouin zone [see Fig.~\ref{fig:ARPES}(b)]. These fits reveal the orbital composition of the Fermi surfaces with states on FS 1 and 2 possessing predominant $d_{xz}$- and $d_{yz}$-character, while states on FS 3 possesses large contributions from the $d_{xy}$-orbital and $d_{xz}/d_{yz}$-orbitals (see SM Sec.~II). These results are qualitatively similar to those obtained earlier for LaOFeAs \cite{Gra09}.

\begin{figure}[!ht]
\begin{center}
\includegraphics[width=8cm]{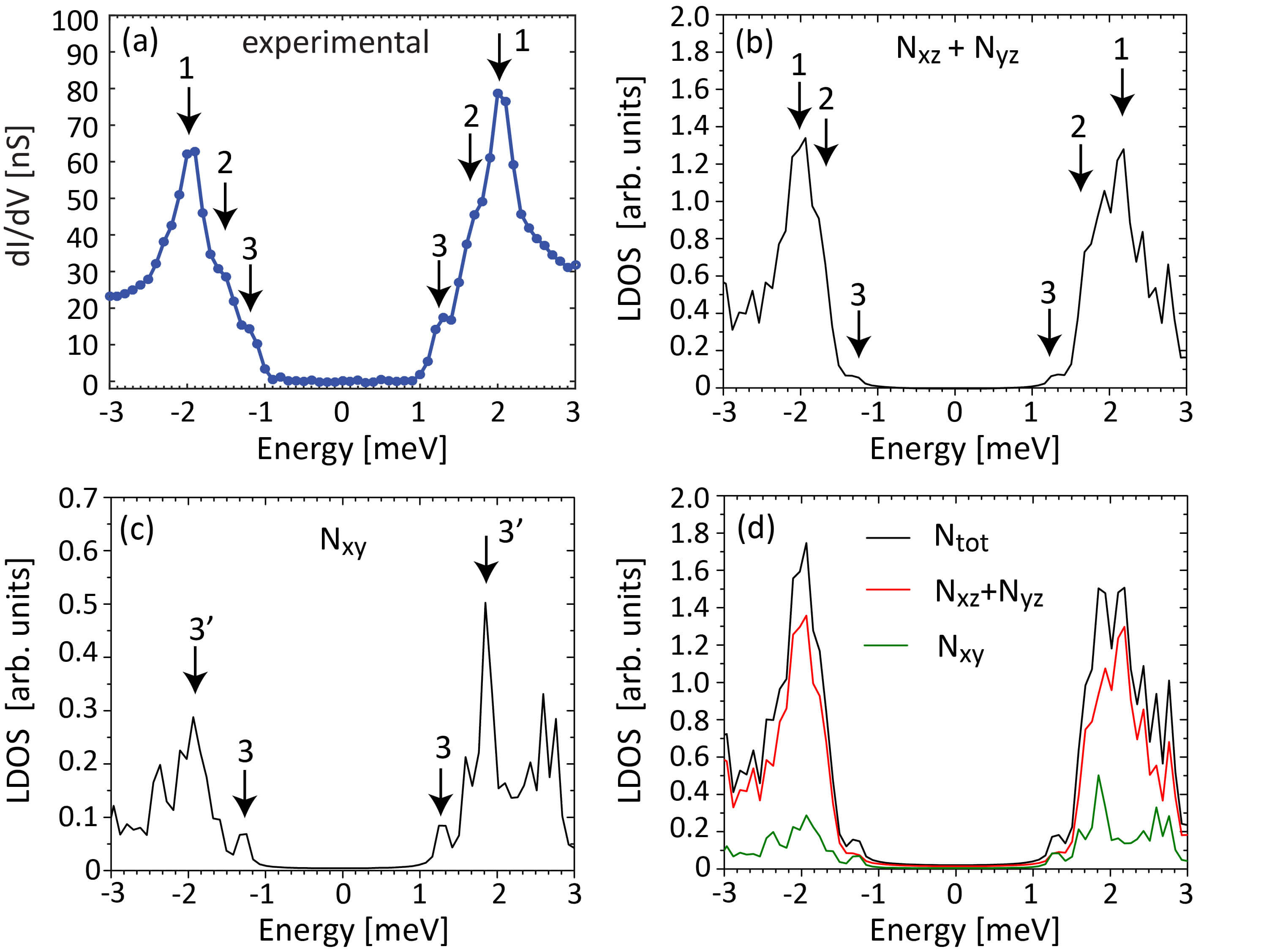}
\end{center}
\caption{(a) Experimental differential tunneling conductance $g({\bf r}, E=eV) = dI({\bf r}, V)/dV$, and theoretical (b) $N_{xz}+N_{yz}$, (c) $N_{xy}$, and (d) $N_{tot}$ in FeSe$_{0.45}$Te$_{0.55}$. The sharp coherence peaks denoted by (1) are associated with the superconducting gap on Fermi surface (1), while the shoulder-like features denoted by (2) and (3) are the smeared-out coherence peaks associated with the superconducting gaps on Fermi surfaces (2) and (3). (3) and (3') are features associated with the maximum and minimum superconducting gaps, $|\Delta_3|^{max}$ and $|\Delta_3|^{min}$, respectively, on Fermi surface 3.  }
\label{fig:SC_LDOS_clean}
\end{figure}

To investigate $dI/dV$ in the superconducting state, we note that the states near the Fermi surfaces consist primarily of contributions from the $d_{xz}$-, $d_{yz}$- and $d_{xy}$-orbitals, and therefore assume that the superconducting order parameter is non-zero for these three orbitals only.
To determine their values in orbital space,  we make use of several characteristic features that occur in $dI/dV$ in the superconducting state of FeSe$_{0.45}$Te$_{0.55}$, as shown in Fig.~\ref{fig:SC_LDOS_clean}(a).
In particular, the $dI/dV$ lineshape possesses a set of easily identifiable coherence peaks [denoted by arrows $1$ in Fig.~\ref{fig:SC_LDOS_clean}(a)], located at $E = \pm 2.0$ meV, and two sets of shoulder-like features [denoted by arrows $2$ and $3$ in Fig.~\ref{fig:SC_LDOS_clean}(a)], at energies $E=\pm 1.8$ meV and $E=\pm 1.2$ meV, respectively. To understand the origin of these features, we plot the theoretically computed and orbitally resolved local density of states (LDOS) $N_{\alpha}({\bf r}, E)$ (see SM Sec.~I) of the $d_{xz}$-orbital ($N_{xz}$) and $d_{yz}$-orbital ($N_{yz}$) in Fig.~\ref{fig:SC_LDOS_clean}(b), and of the $d_{xy}$-orbital ($N_{xy}$) in Fig.~\ref{fig:SC_LDOS_clean}(c). A plot of the total LDOS, $N_{tot}$, i.e., the sum over all orbital contributions, shown in Fig.\ref{fig:SC_LDOS_clean}(d), reveals that the largest contribution to $N_{tot}$, arises from the $d_{xz}$- and $d_{yz}$-orbitals. Moreover, $N_{xz}$ and $N_{yz}$ [Fig.~\ref{fig:SC_LDOS_clean}(b)] exhibit the same features seen experimentally: a set of coherence peaks denoted by $1$, and two shoulder like features at lower energy denoted by $2$ and $3$. To match the energy position of these features to those observed experimentally, we take the superconducting gaps in orbital space to be given by $\Delta_{11}=\Delta_{22}=0.55$ meV, and $\Delta_{44}=0.38$ meV. A plot of the superconducting gaps along the three Fermi surface sheets, denoted by $\Delta^{{\rm FS}}_{1,2,3}$, shown in Fig.~\ref{fig:ARPES}(c), allows us to identify the origin of these characteristic features.  In particular, the largest superconducting gap, exhibiting only a very weak variation with angle, is found on Fermi surface 1 with $\Delta^{max}_1 \approx 2$ meV and gives rise to the coherence peaks denoted by arrows $1$ in Fig.\ref{fig:SC_LDOS_clean}(b). The superconducting gap on Fermi surface 2, $\Delta^{{\rm FS}}_2({\bf k})$, exhibits a weak anisotropy and varies between $\Delta_2^{min}=1.67$ meV and  $\Delta_2^{max}=1.73$ meV, leading to the shoulder-like feature in the LDOS indicated by arrows $2$ in Fig.~\ref{fig:SC_LDOS_clean}(b); this feature can be interpreted as broadened coherence peaks. Finally, the gap on Fermi surface 3 possesses the largest anisotropy varying between $|\Delta_3|^{min}=1.2$ meV and $|\Delta_3|^{max}=1.8$ meV, with $|\Delta_3|^{min}$ determining the position of the shoulder-like feature denoted by arrows $3$. This strong anisotropy is a direct result of the varying orbital composition of states along FS 3, and a superconducting gap in the $d_{xy}$-orbital, that is significantly smaller than that in the $d_{xz}$- and $d_{yz}$-orbitals (for a detailed discussion, see SM Sec.~II). Moreover, as $|\Delta_3|^{max}$ is quite close to $\Delta^{max}_2$, it is impossible to resolve its signature in $N_{xz}$ and $N_{yz}$ from that of $\Delta^{{\rm FS}}_2({\bf k})$. However, the signatures associated with $|\Delta_3|^{min}$ and $|\Delta_3|^{max}$ can be clearly identified in $N_{xy}$ [Fig.\ref{fig:SC_LDOS_clean}(c)] as indicated by arrows $3$ and $3'$.
While our model assumes only two independent values of the superconducting gaps, the energies of all three features in the theoretically computed LDOS -- the coherence peaks and two shoulder-like features -- are in very good agreement with the experimental findings, providing strong evidence for the validity of the extracted bandstructure and superconducting gaps.

The above results shed some light on the ongoing controversy regarding the size and anisotropy of the superconducting gaps reported by angle-resolved photoemission experiments on Fe$_{1.03}$Te$_{0.7}$Se$_{0.3}$ \cite{Nak10}, FeTe$_{0.6}$Se$_{0.4}$ \cite{Oka12} and FeTe$_{0.55}$Se$_{0.45}$ \cite{Miao12}. These experiments have reported not only significantly different values of the maximum superconducting gaps, varying between 2 meV  \cite{Oka12} and 4 meV \cite{Nak10,Miao12} (with the latter being inconsistent with our results), but also disagree on whether the gaps are isotropic \cite{Nak10,Miao12} or highly anisotropic \cite{Oka12}. In particular, the report of an isotropic superconducting gap on the electron-like Fermi surface 3 is not only inconsistent with our findings, but also with those of angle-resolved specific heat measurements \cite{Zeng10}. The latter also reported a gap minimum on FS 3 along the $\Gamma-M$ direction, in agreement with our findings.

\begin{figure}[!ht]
\begin{center}
\includegraphics[width=8cm]{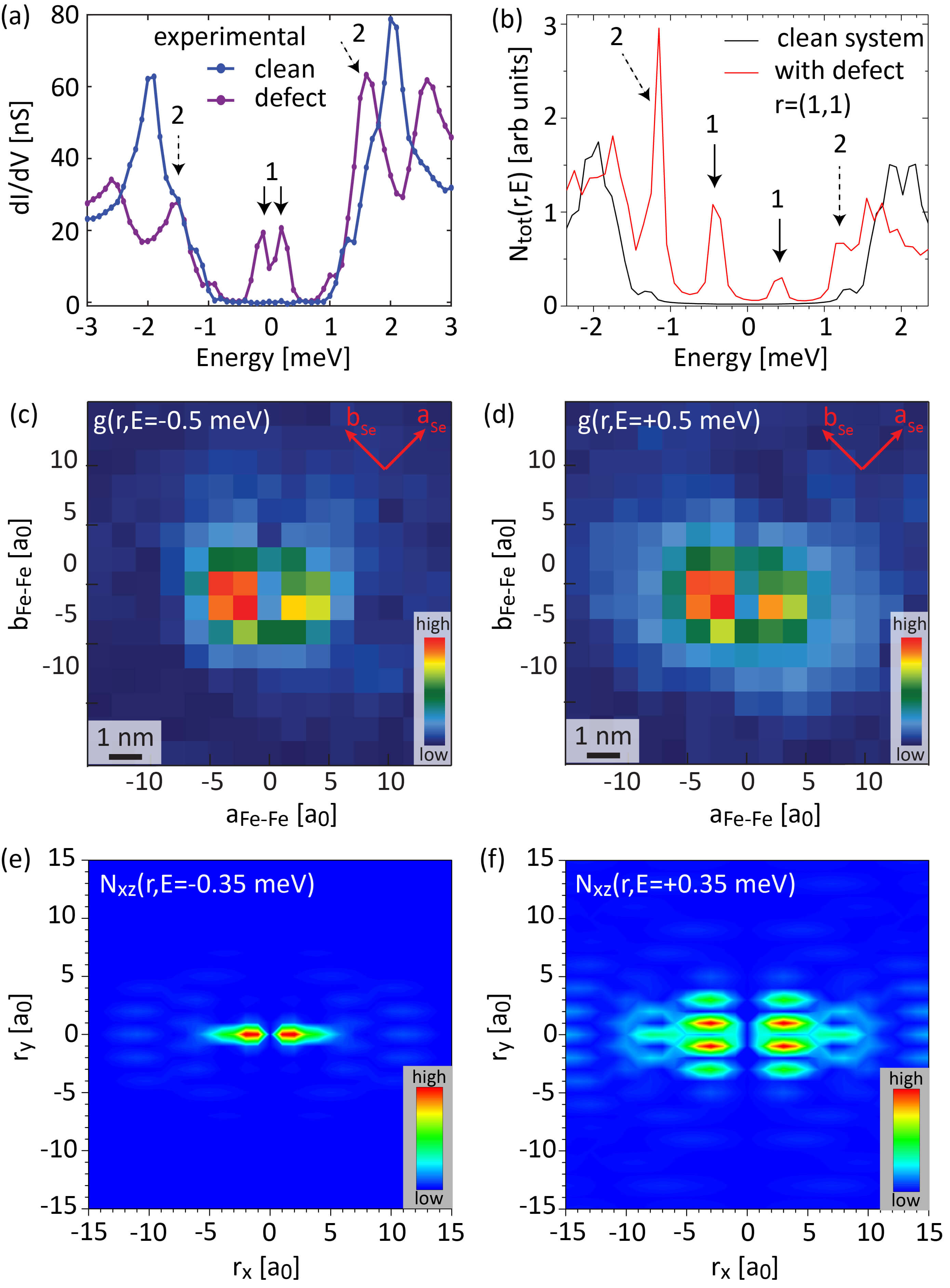}
\end{center}
\caption{(a) Experimental $dI/dV({\bf r}, E)$ and (b) theoretical $N_{tot}({\bf r},E)$ at ${\bf r}=(1,1)$ with the defect located at ${\bf R}=(0,0)$. Experimental $dI/dV({\bf r}, E)$ at the (c) particle-like, and (d) hole-like branch of the impurity bound state. Experimental $dI/dV({\bf r}, E)$ was recorded at 280 mK using a modulation of 100 $\mu$V, setup bias of -20 mV, and tunneling current of 500 pA. Theoretical $N_{xz}({\bf r},E)$ at the bound state energies: (e) $E_b=-0.35$ meV, (f) $E_b=+0.35$ meV. }
\label{fig:exp_defects}
\end{figure}

Further important insight into the electronic structure of FeSe$_{0.45}$Te$_{0.55}$ can be gained by considering the experimentally measured $dI/dV$ near defects, as shown in Fig.~\ref{fig:exp_defects}(a). Due to the unconventional symmetry of the superconducting order parameter, magnetic as well as non-magnetic defects lead to the emergence of impurity states inside the superconducting gap \cite{Gas13}. These impurity states possess a dumbbell-like spatial structure [Fig.~\ref{fig:exp_defects}(c) and (d)], which breaks the electronic $C_4$-symmetry of the system, and show little difference between the particle- ($E<0$) and hole-like ($E>0$) branches of the defect state. The proximity to the nematic phase suggests that this dumbbell-like structure might arise from the pinning of dynamic nematic fluctuations by the defects, similar to the pinning of dynamic charge- or spin-density wave fluctuations \cite{Nyb08}. This pinning effectively increases the spatial extent of a defect's scattering potential along the $x$- or $y$-axis. Alternatively, the observed broken $C_4$-symmetry of the defect states could reflect the presence of orbital selective Mottness \cite{Sprau16a,Sprau16b,Yin11,Med14}.

Here, we consider the former possibility, and investigate the combined effects of defects and pinned nematic fluctuations on the electronic structure of FeSe$_{0.45}$Te$_{0.55}$ by employing the Hamiltonian
\begin{align}
\label{eq:Ham_scat}
 H_{scat} =&  U_0 \sum_{\sigma}\sum_{\alpha=1}^{5} c_{{\bf R},\alpha,\sigma}^{\dagger} c_{{\bf R},\alpha,\sigma} + U_p \Phi ({\bf R}) \nonumber \\
 & + g \sum_{{\bf r},\alpha, \beta, \sigma}\Phi \left( {\bf r} \right) c^\dagger_{{\bf r},\alpha,\sigma} c_{{\bf r},\beta,\sigma} \ .
\end{align}
where the first term describes a non-magnetic defect located at ${\bf R}=(0,0)$ that leads to electronic on-site, intra-orbital scattering only, the second term describes the defect-induced pinning of nematic fluctuations represented by the field $\Phi ({\bf R})$, and the last term represents the interaction between the nematic fluctuations and the conduction electrons with interaction strength $g$. The pinning of nematic fluctuations induces additional scattering potentials for the conduction electrons, described by  $U_{\bf r} = g \langle \Phi ({\bf r}) \rangle = - g U_p \chi_n({\bf r - R}, \omega=0)$, where  $\chi_n$ is the susceptibility of the nematic fluctuations. In general, the pinning of the nematic fluctuations could also affect electronic hopping elements, which will not be considered here. The spatial extent and strengths of these pinning induced scattering potentials is in general determined by the correlation lengths of the nematic fluctuations, as reflected in $\chi_n$. While the calculation of the latter is beyond the scope of this article, we make use of the fact that it is highly directional \cite{Fer12,Fer12a,Kar15,Wang15}. We therefore model the pinning of nematic fluctuations as leading to scattering potentials $U_1 =g \left\langle \Phi \left( {\bf R} \pm {\hat x} \right) \right\rangle$ and $U_2 =g \left\langle \Phi \left( {\bf R} \pm 2{\hat x} \right) \right\rangle$ along the $x$-axis only. For concreteness, we use $U_0=100$ meV, $U_1 =75$ meV, and $U_2 =50$ meV for the defects considered below.

The resulting $N_{tot}$ (see SM Sec.~I) near the defect site, shown in Fig.~\ref{fig:exp_defects}(b), reveals as expected impurity states inside the superconducting gap located at $E_b=\pm 0.35$ meV denoted by arrows $1$. In addition, $N_{tot}$ shows a strong enhancement close to the edge of the superconducting gap [denoted by arrows $2$]. Both of these features are in agreement with those observed experimentally [see corresponding arrows in Fig.~\ref{fig:exp_defects}(a)]. A comparison of the spatially and orbitally resolved LDOS (see Fig.S2 in SM Sec.~III) shows that the $d_{xz}$-orbital possesses the largest LDOS at $E_b$ (for our choice of $x$- and $y$-axes and direction of nematic fluctuations). It is therefore likely that the largest contribution to the experimentally measured $dI/dV$ at the bound state energies arises from tunneling into the $d_{xz}$-orbitals. We therefore present in Figs.~\ref{fig:exp_defects}(e) and (f) the spatially resolved $N_{xz}$ for the particle- ($E=-0.35$ meV) and hole-like ($E=+0.35$ meV) branches of the defect state. Due to the pinning of the nematic fluctuations, the bound state extends primarily along the $x$-axis, and possesses a dumbbell-like shape, similar to the one observed experimentally in Figs.~\ref{fig:exp_defects}(c) and (d). Note that the general spatial structure of the bound state changes only slightly between the particle- and hole-like branches, in agreement with the experimental observation, and in contrast to the characteristic $45^\circ$ rotation observed in the cuprate \cite{Hud01} and heavy fermion superconductors \cite{Zhou13,Dyke15}. The reason for this striking difference lies in the orbital structure of FeSe$_{0.45}$Te$_{0.55}$: as the main contribution to the defect state resides in the $d_{xz}$-orbital, the orbital's spatial structure restricts the bound state to predominantly lie along the $x$-axis, and thus does not allow for a rotation of the spatial bound state pattern. Note that for a tetragonal system, nematic fluctuations both along the $x$- and $y$-axes are allowed, and the orientation of the dumbbells will therefore vary between the domain of nematic fluctuations they reside in.

 \begin{figure}[!ht]
\begin{center}
\includegraphics[width=8cm]{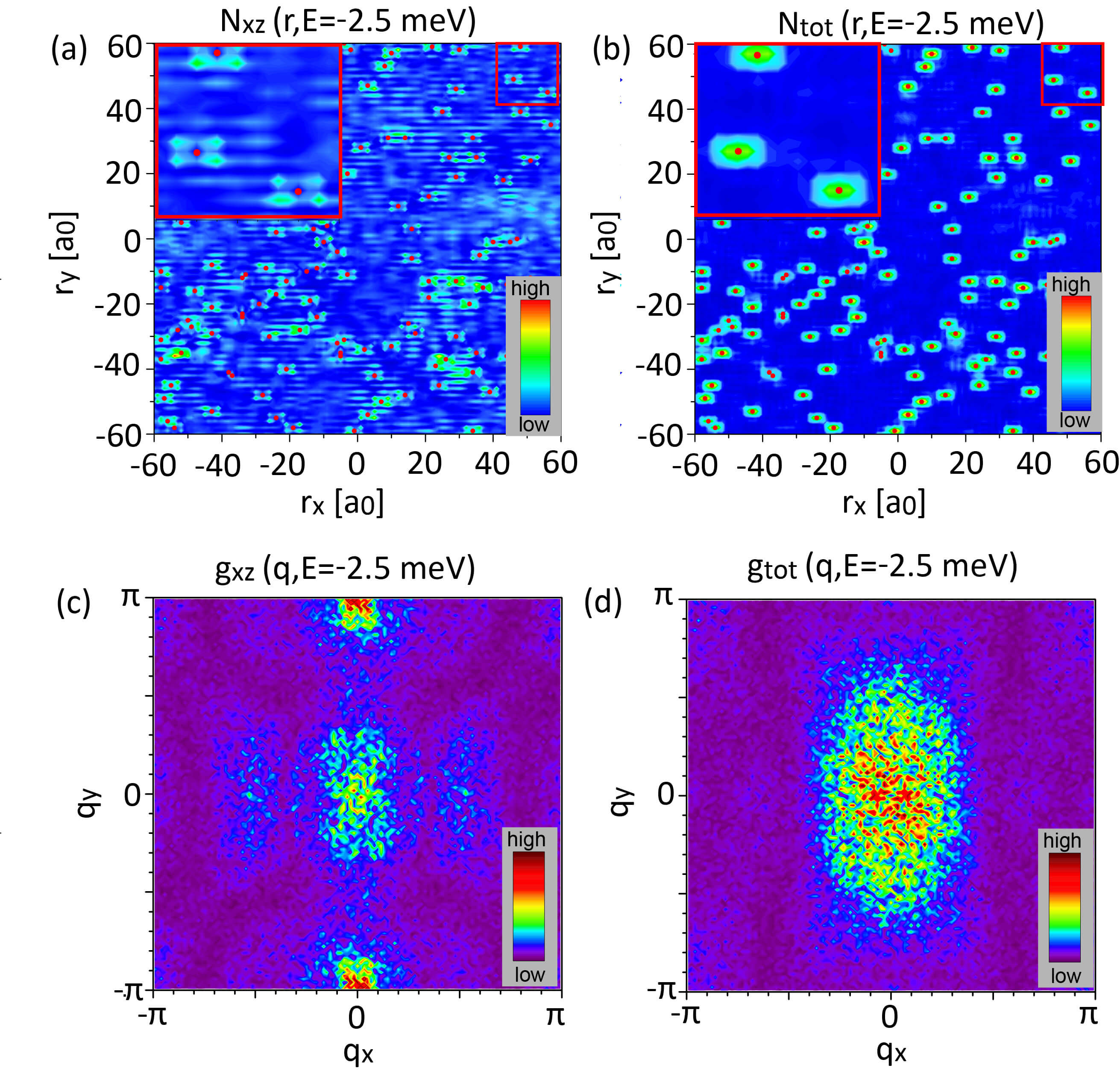}
\end{center}
\caption{Disordered system with a concentration of 1\% of defects pinning nematic fluctuations and giving rise to the same scattering potentials as the defect in Fig.~\ref{fig:exp_defects}(b). Resulting theoretical (a) $N_{xz}({\bf r},E)$ , (b) $N_{tot}({\bf r},E)$, (c) $g_{xz}({\bf q},E)$ , (b) $g_{tot}({\bf q},E)$ at $E=-2.5$ meV.}
\label{fig:QPI}
\end{figure}
Recent STM experiments \cite{Sin15} reported a breaking of the electronic $C_4$-symmetry, and an approximate $C_2$-symmetry of the Fourier transformed differential conductance, $g({\bf q},E)$, outside the nematic phase in doped FeSe$_{0.4}$Te$_{0.6}$.  To investigate the origin of this broken $C_4$-symmetry, we consider a system containing 1\% of defects that pin nematic fluctuations and give rise to the same scattering potentials as the defect in Fig.~\ref{fig:exp_defects}(b) [the defects are indicated by red dots in Fig.~\ref{fig:QPI}(a) and (b)]. We assume that the correlation length associated with the pinned nematic fluctuations is larger than the inter-defect distance, such that all pinned fluctuations are aligned in the same direction. In Figs.~\ref{fig:QPI}(a) and (b), we plot the resulting $N_{xz}({\bf r},E)$ and $N_{tot}({\bf r},E)$, respectively, outside the superconducting gap at $E=-2.5$ meV. The zoom-in in the upper left corner of the area denoted by a red square reveals an approximate $C_2$-symmetry around the defects. Moreover, the defects give rise to strong spatial oscillations in $N_{xz}({\bf r},E)$, possessing a wave-length of $\lambda \approx 2 a_0$. By Fourier-transforming $N_{xz}({\bf r},E)$ and $N_{tot}({\bf r},E)$, we obtain $g_{xz}({\bf q},E)$ and $g_{tot}({\bf q},E)$ shown in Fig.~\ref{fig:QPI}(c) and (d) respectively. As expected, they possess an approximate $C_2$-symmetry, and exhibit a structure that is very similar to the one observed experimentally in $g({\bf q},E)$ by Singh {\it et al.}~\cite{Sin15} (see SI Sec.~IV). We thus conclude that even outside the nematic phase, the pinning of strong nematic fluctuations can break the electronic $C_4$-symmetry of the system. The theoretical $g_{xz}({\bf q},E)$ also provide insight into the origin of the strong spatial oscillations in the LDOS. The large intensity in $g_{xz}({\bf q},E)$ at small wave-vectors around $(0,0)$ arises from scattering within the three Fermi surfaces, whereas the large intensity around $(0,\pm \pi)$ [see Fig.~\ref{fig:QPI}(c)] arises from scattering between the electron-like Fermi surface sheets closed around $(0, \pm \pi)$ and the two hole-like Fermi surfaces closed around $(0,0)$ [see Fig.~\ref{fig:ARPES}(b)]. Thus the $\lambda \approx 2 a_0$ oscillations in the LDOS are a direct signature of the interband scattering induced by defects.

In conclusion, we have shown that the differential conductance, $dI/dV$, measured in the doped iron chalcogenide FeSe$_{0.45}$Te$_{0.55}$ provides important insight into the structure of the superconducting order parameter on three Fermi surface sheets. We extracted the magnitude and orbital content of the superconducting gaps using characteristic features in the $dI/dV$ lineshape. This allowed us to obtain nearly isotropic superconducting gaps on the two hole-like Fermi surfaces, and strongly anisotropic gap on the electron-like Fermi surface. Moreover, we demonstrated that the correlated pinning of nematic fluctuations by defects can explain not only the dumbbell-like shape of the induced impurity states, but also the broken $C_4$-symmetry of the observed $g({\bf q,E})$ for energies outside the superconducting gap. Further insight into the role played by orbital-selective Mottness into determining the system's electronic structure, and response to impurities, could be provided by future experiments utilizing the yet unexplored dispersive features in the quasi-particle interference.

\begin{acknowledgments}
We would like to thank P. Hirschfeld for stimulating discussions, and P. Wahl for providing us with Figs.S3A and C.  This work was supported by the U. S. Department of Energy, Office of Science, Basic Energy Sciences, under Award No. DE-FG02-05ER46225 (DKM). Experimental studies, as well as the work by SS and JVD are supported by the Center for Emergent Superconductivity, an Energy Frontier Research Center, headquartered at Brookhaven National Laboratory and funded by the U.S. Department of Energy, under DE-2009-BNL-PM015.
\end{acknowledgments}

\end{document}


\fontsize{11}{13}

\begin{center}
{\large {\bf  Supplemental Online Material for}} \\[0.5cm]
\end{center}

\title{Orbital superconductivity, defects and pinned nematic fluctuations in the doped iron chalcogenide FeSe$_{0.45}$Te$_{0.55}$}
\author{Saheli Sarkar$^1$}
\author{John Van Dyke$^{1,2}$}
\author{Peter Sprau$^{3,4}$}
\author{Freek Massee$^{3,4,5}$}
\author{Ulrich Welp$^6$}
\author{Wai-Kwong Kwok$^6$}
\author{J.C. Seamus Davis$^{3,4,7,8}$}
\author{Dirk K. Morr$^1$}

\affiliation{$^1$ University of Illinois at Chicago, Chicago, IL 60607, USA}

\affiliation{$^2$ Department of Physics and Astronomy, Iowa State University, Ames, Iowa 50011, USA}

\affiliation{$^3$ LASSP, Department of Physics, Cornell University, Ithaca, NY 14853, USA}

\affiliation{$^4$ CMPMS Department, Brookhaven National Laboratory, Upton, NY 11973, USA}

\affiliation{$^5$ Laboratoire de Physique des Solides (CNRS UMR 8502), B\^{a}timent 510, Universit\'{e} Paris-Sud/Universit\'{e} Paris-Saclay, 91405 Orsay, France}

\affiliation{$^6$  Materials Science Division Argonne National Laboratory, Argonne, IL 60439, USA}

\affiliation{$^7$ School of Physics and Astronomy, University of St. Andrews, Fife KY16 9SS, Scotland, UK }

\affiliation{$^8$ Tyndall National Institute, University College Cork, Cork T12R5C, Ireland}

\date{\today}

\maketitle

\section{Theoretical Formalism}
\label{sec:theory}

In this article we use a five-orbital model introduced by Graser \textit{et al.} \cite{Gra09} to describe the electronic structure of FeSe$_{0.45}$Te$_{0.55}$ in the normal and superconducting state. The orbitals involved in this model are the five Fe $3d$ orbitals, which are denoted using the following convention: $1$ corresponds to the $d_{xz}$-orbital, $2$ corresponds to the $d_{yz}$-orbital, $3$ corresponds to the $d_{x^{2}-y^{2}}$-orbital, $4$ corresponds to the $d_{xy}$-orbital and $5$ corresponds to the $d_{3z^{2}-r^{2}}$-orbital. The Hamiltonian for the normal state of FeSe$_{0.45}$Te$_{0.55}$ is then given by
\begin{equation} \label{eq:Ham_k}
 H_{0} = \sum_{k,\sigma}\sum_{\alpha,\beta=1}^{5}\varepsilon_{k}^{\alpha \beta}c_{k,\alpha,\sigma}^{\dagger} c_{k,\beta,\sigma} \ ,
\end{equation}
where $\alpha$,$\beta$ are the orbital indices as described above. Here, $\varepsilon_{k}^{\alpha \beta}$ with $\alpha \neq \beta$ ($\alpha =\beta$) describe the parts of the dispersions that arise from inter-orbital (intra-orbital) hopping, with $c_{k,\alpha,\sigma}^{\dagger} (c_{k,\beta,\sigma})$ creating (annihilating) an electron in orbital $\alpha$ with momentum {\bf k} and spin $\sigma$. All orbitals and hopping elements are defined in the 1Fe Brillouin zone. In what follows, we employ the following form of $\varepsilon_{k}^{\alpha \beta}$  \cite{Gra09}
\begin{align}
\varepsilon_{11} & =2t_{x}^{11}\cos\textit{k}_{x} + 2t_{y}^{11}\cos\textit{k}_{y} + 4t_{xy}^{11}\cos\textit{k}_{x}\cos\textit{k}_{y} + 2t_{xx}^{11}(\cos2\textit{k}_{x} - \cos2\textit{k}_{y}) \nonumber \\
  & \quad + 4t_{xxy}^{11}\cos2\textit{k}_{x}\cos2\textit{k}_{y} + 4t_{xyy}^{11}\cos2\textit{k}_{y}\cos\textit{k}_{x}
+ 4t_{xxyy}^{11}\cos(2\textit{k}_{x})\cos(2\textit{k}_{y})\nonumber  \\
\varepsilon_{22} & =2t_{y}^{11}\cos\textit{k}_{x} + 2t_{x}^{11}\cos\textit{k}_{y} + 4t_{xy}^{11}\cos\textit{k}_{x}\cos\textit{k}_{y} - 2t_{xx}^{11}(\cos2\textit{k}_{x} - \cos2\textit{k}_{y}) \nonumber \\
 & \quad + 4t_{xyy}^{11}\cos2\textit{k}_{x}\cos2\textit{k}_{y} + 4t_{xxy}^{11}\cos2\textit{k}_{y}\cos\textit{k}_{x}
+ 4t_{xxyy}^{11}\cos(2\textit{k}_{x})\cos(2\textit{k}_{y}) \nonumber  \\
\varepsilon_{33} &=2t_{x}^{33}(\cos\textit{k}_{x} + \cos\textit{k}_{y}) + 4t_{xy}^{33}\cos\textit{k}_{x}\cos\textit{k}_{y} + 2t_{xx}^{33}(\cos2\textit{k}_{x} + \cos2\textit{k}_{y}) \nonumber  \\
\varepsilon_{44} & =2t_{x}^{44}(\cos\textit{k}_{x} + \cos\textit{k}_{y}) + 4t_{xy}^{44}\cos\textit{k}_{x}\cos\textit{k}_{y} + 2t_{xx}^{44}(\cos2\textit{k}_{x} + \cos2\textit{k}_{y}) \nonumber \\
& \quad +4t_{xxy}^{44}(\cos2\textit{k}_{x}\cos\textit{k}_{y} + \cos2\textit{k}_{y}\cos\textit{k}_{x}) + 2t_{xxyy}^{44}\cos2\textit{k}_{x}\cos2\textit{k}_{y} \nonumber \\
\varepsilon_{55}&=2t_{x}^{55}(\cos\textit{k}_{x} + \cos\textit{k}_{y}) + 2t_{xx}^{55}(\cos2\textit{k}_{x} + \cos2\textit{k}_{y}) + 4t_{xxy}^{55}(\cos2\textit{k}_{x}\cos\textit{k}_{y} + \cos2\textit{k}_{y}\cos\textit{k}_{x})
\nonumber \\
&\quad +4t_{xxyy}^{55}\cos2\textit{k}_{x}\cos2\textit{k}_{y} \nonumber  \\
\varepsilon_{12} &= -4t_{xy}^{12}\sin\textit{k}_{x}\sin\textit{k}_{y} - 4t_{xxy}^{12}(\sin2\textit{k}_{x}\sin\textit{k}_{y}) - 4t_{xxyy}^{12}\sin2\textit{k}_{x}\sin2\textit{k}_{y} \nonumber  \\
\varepsilon_{13}& = 2it_{x}^{13}\sin\textit{k}_{y} + 4it_{xy}^{13}\sin\textit{k}_{y}\cos\textit{k}_{x} - 4it_{xxy}^{13}(\sin2\textit{k}_{y}\cos\textit{k}_{x} - \cos2\textit{k}_{x}\sin\textit{k}_{y}) \nonumber  \\
\varepsilon_{23} &= -2it_{x}^{13}\sin\textit{k}_{x} - 4it_{xy}^{13}\sin\textit{k}_{x}\cos\textit{k}_{y} + 4it_{xxy}^{13}(\sin2\textit{k}_{x}\cos\textit{k}_{y} - \cos2\textit{k}_{y}\sin\textit{k}_{x}) \nonumber  \\
\varepsilon_{14}&=2it_{x}^{14}\sin\textit{k}_{x} + 4it_{xy}^{14}\cos\textit{k}_{y}\sin\textit{k}_{x} + 4it_{xxy}^{14}\sin2\textit{k}_{x}\cos\textit{k}_{y} \nonumber  \\
\varepsilon_{24}&=2it_{x}^{14}\sin\textit{k}_{y} + 4it_{xy}^{14}\cos\textit{k}_{x}\sin\textit{k}_{y} + 4it_{xxy}^{14}\sin2\textit{k}_{y}\cos\textit{k}_{x} \nonumber  \\
\varepsilon_{15}&=2it_{x}^{15}\sin\textit{k}_{y} - 4it_{xy}^{15}\sin\textit{k}_{y}\cos\textit{k}_{x} - 4it_{xxyy}^{15}\sin2\textit{k}_{y}\cos2\textit{k}_{x} \nonumber  \\
\varepsilon_{25}&=2it_{x}^{15}\sin\textit{k}_{x} - 4it_{xy}^{15}\sin\textit{k}_{x}\cos\textit{k}_{y} - 4it_{xxyy}^{15}\sin2\textit{k}_{x}\cos2\textit{k}_{y} \nonumber  \\
\varepsilon_{34} &= 4t_{xxy}^{34}(\sin2\textit{k}_{y}\sin\textit{k}_{x} - \sin2\textit{k}_{x}\sin\textit{k}_{y}) \nonumber \\
\varepsilon_{35} &= 2t_{x}^{35}(\cos\textit{k}_{x} - \cos\textit{k}_{y}) + 4t_{xxy}^{35}(\cos2\textit{k}_{x}\cos\textit{k}_{y} - \cos2\textit{k}_{y}\cos\textit{k}_{x}) \nonumber \\
\varepsilon_{45} &= 4t_{xy}^{45}\sin\textit{k}_{x}\sin\textit{k}_{y} + 4t_{xxyy}^{45}\sin2\textit{k}_{x}\sin2\textit{k}_{y}
\end{align}
\newpage

The hopping elements were extracted from a fit of the theoretical dispersion arising from the Hamiltonian of Eq.(\ref{eq:Ham_k}) to the electronic dispersion measured in angle-resolved photoemission spectroscopy (ARPES) experiments \cite{Tam10}, as shown in Fig.~1(a) of the main text. They are given by (all hopping elements are given in units of meV):
\begin{table}[h!]
\centering
\caption{Intra-orbital Hopping parameters used }
\label{table:1}
\vspace{3mm}
\begin{tabular}{||c | c c c c c c c||}
 \hline
 Orbitals &$t_{x}^{pp}$ & $t_{y}^{pp}$ & $t_{xy}^{pp}$ & $t_{xx}^{pp}$ & $t_{xxy}^{pp}$ & $t_{xyy}^{pp}$ & $t_{xxyy}^{pp}$ \\ [0.5ex]
 \hline\hline
 pp=11 & -11.0 & -43.0 & 28.0 & 2.0 & -3.5 & 0.5 & 3.5 \\
 pp=33 & 32.0 &  & -10.5 & -2.0 &  &  &  \\
 pp=44 & 22.0 &  & 15.0 & -2.0 & -3.0 &  & -3.0 \\
 pp=55 & -10.0 &  &  & -4.0 & 2.0 &  & -1.0 \\ [1ex]
 \hline
\end{tabular}
\end{table}
\begin{table}[h!]
\centering
\caption{Inter-orbital Hopping parameters used }
\label{table:2}
\vspace{3mm}
\begin{tabular}{||c | c c c c ||}
 \hline
 Orbitals & $t_{x}^{pq}$ & $t_{xy}^{pq}$ & $t_{xxy}^{pq}$ & $t_{xxyy}^{pq}$ \\ [0.5ex]
 \hline\hline
 pq=12 &  & 5.0 & -1.5 & 3.5 \\
 pq=13 & -35.4 & 9.9 & 2.1 &  \\
 pq=14 & 33.9 & 1.4 & 2.8 &  \\
 pq=15 & -19.8 & -8.5 &  & -1.4 \\
 pq=34 &  &  & -1.0 &  \\
 pq=35 & -30.0 &  & -5.0 &  \\
 pq=45 &  & -15.0 &  & 1.0 \\ [1ex]
  \hline
\end{tabular}
\end{table}
The onsite energies for orbitals are given by  $\varepsilon_{1}$= 7.0 meV, $\varepsilon_{2}$=7.0meV, $\varepsilon_{3}$=-25.0meV,
$\varepsilon_{4}$=20.0meV, and $\varepsilon_{5}$=-25.1 meV.
In real space, the Hamiltonian is given by
\begin{equation}\label{hk3}
H_{0} = \sum_{i,j,\sigma}\sum_{\alpha,\beta=1}^{5}(t_{i,j}^{\alpha,\beta}c_{i,\alpha,\sigma}^{\dagger}c_{j,\beta,\sigma} + H.c.)
\end{equation}
Moreover, the superconducting pairing is described by the mean-field Hamiltonian
\begin{align}
H_{SC}^{MF} & = \sum_{\langle {\bf r},{\bf r'} \rangle }\sum_{\alpha=1}^{5} \Delta_{\alpha \alpha} c_{{\bf r},\alpha,\uparrow}^{\dagger} c_{{\bf r'},\alpha,\downarrow}^{\dagger} + H.c.
\end{align}
where the first sum only runs over next-nearest neighbor Fe sites in the 1 Fe unit cell, giving rise to a superconducting order parameter with $s_\pm$-symmetry.
In addition, we consider the scattering off non-magnetic defects described by
\begin{align}
H_{scat} =  \sum_{{\bf R},\sigma}\sum_{\alpha=1}^{5} U_{\bf R} c_{{\bf R},\alpha,\sigma}^{\dagger} c_{{\bf R},\alpha,\sigma}  \ .
\label{eq:Ham_scatt}
\end{align}
where the sum runs over all defect sites ${\bf R}$. The total Hamiltonian in real space in the superconducting state is then given by $H=H_0+H_{SC}^{MF}+H_{scat}$.

To compute the orbitally, spatially, and energy-resolved local density of state, it is necessary to compute the associated Green's functions in real space. To this end, we employ the Nambu formalism, and first define a five-orbital spinor
\begin{equation}\label{hk8}
\Psi^{\dagger} = (\ldots,c_{j,1,\uparrow}^{\dagger},c_{j,1,\downarrow},c_{j,2,\uparrow}^{\dagger},c_{j,2,\downarrow},c_{j,3,\uparrow}^{\dagger},c_{j,3,\downarrow},c_{j,4,\uparrow}^{\dagger},c_{j,4,\downarrow},c_{j,5,\uparrow}^{\dagger},c_{j,5,\downarrow},\ldots)
\end{equation}
where the index $j$ represents the $j'th$ site in the system, the second index $1, ..., 5$ represents the orbital, and the last index represents the spin. Using this spinor, we can write the total Hamiltonian
\begin{equation}
H =\Psi_{\sigma}^{\dagger}\hat{H}\Psi_{\sigma}
\end{equation}
with $\hat{H}$ being the Hamiltonian matrix in real, orbital, and spin space.

The Green's function matrix in Nambu and Matsubara space is then defined via
\begin{equation}
\hat{G}(\tau) = - \langle T_{\tau}\Psi_{\vec{\bf r}}(\tau)\Psi_{\vec{\bf r}'}^{\dagger}(0)\rangle
\end{equation}
and the retarded Green's function is then given by
\begin{equation}
\hat{G}^r(E) =  \left[ (E + i \delta)  \hat{I} - \hat{H} \right]^{-1}
\end{equation}
where $\hat{I}$ is the identity matrix, and $\delta = 0^+$. The local density of states at site ${\bf r}$ and orbital $\alpha$ is then obtained via
\begin{align}
N_\alpha({\bf r},E) = -\frac{1}{\pi} {\rm Im} G^r_{\alpha\alpha}({\bf r},{\bf r},E)
\end{align}
where $G^r_{\alpha\alpha}({\bf r},{\bf r},E)$ is the element of $\hat{G}^r(E)$ that represents the local in orbital and real space Green's function for orbital $\alpha$ and site ${\bf r}$. The total density of states is then given by $N_{tot}({\bf r},E) = \sum_{\alpha=1}^5 N_\alpha({\bf r},E)$.\\

Finally, we denote by $g_\alpha({\bf q},E)$ the Fourier transform into momentum space of the density of states $N_\alpha({\bf r},E)$ of orbital $\alpha$. Correspondingly, $g_{tot}({\bf q},E)$ is the Fourier transform of $N_{tot}({\bf r},E)$.

\section{Orbital content of states on the Fermi surface}

\begin{figure}[!ht]
\renewcommand{\thefigure}{S\arabic{figure}}
\begin{center}
\includegraphics[width=12cm]{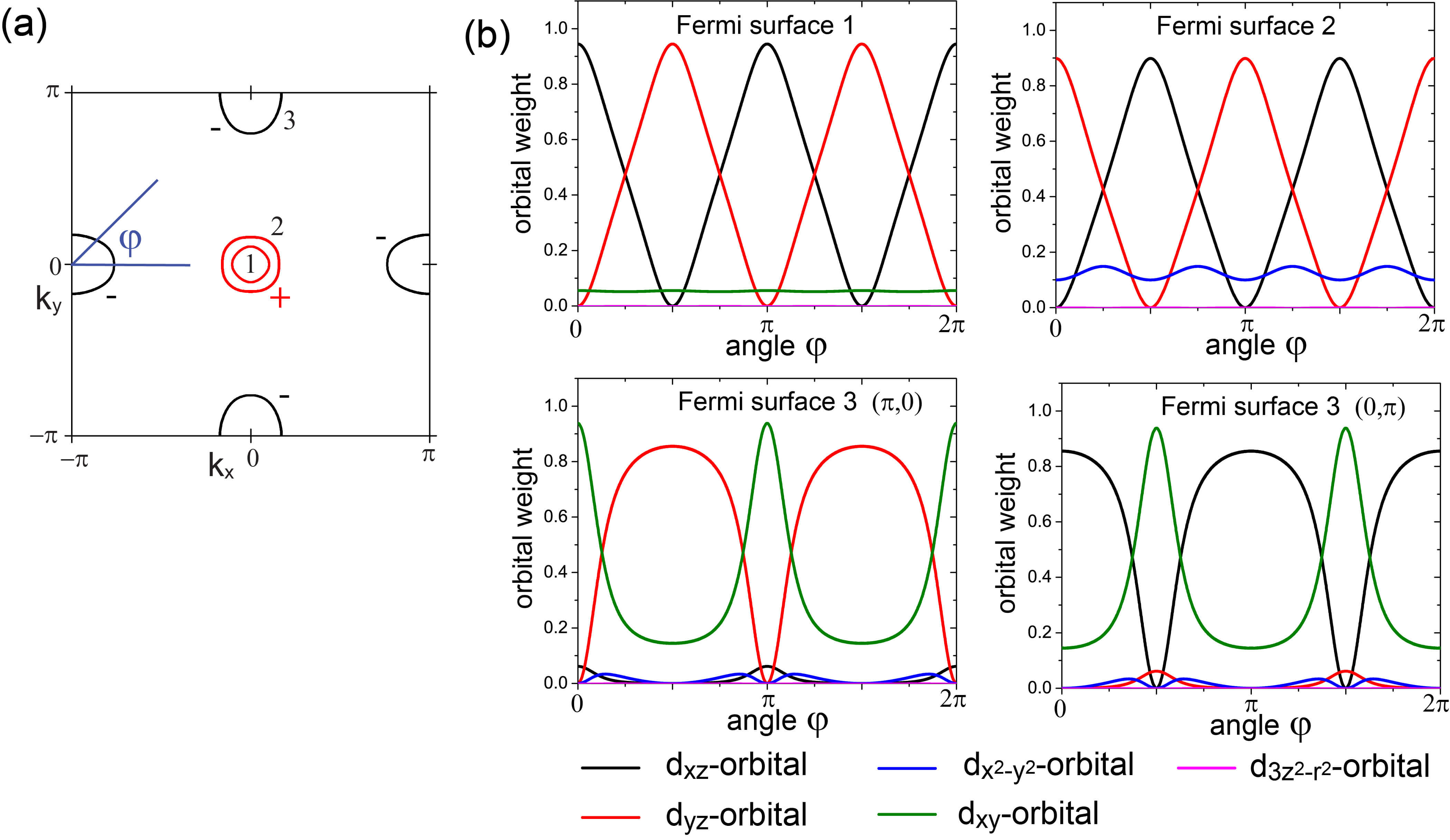}
\caption{(a) Plot of the three Fermi surface sheets with two hole-like FS pockets around $\Gamma$ [$(0,0)$] point and one electron-like FS pocket around $M$ [$(\pm \pi,0), (0,\pm \pi)$] points in the 1Fe BZ. "$\pm$" indicate the phase of the superconducting $s_\pm$-wave order parameter. (b) Angular dependence of the orbital weight (total weight at each momentum point is unity) around the three Fermi surface sheets.
}
\label{fig:orb}
\end{center}
\end{figure}

By fitting the bandstructure of the five-orbital, tight-binding Hamiltonian \cite{Gra09} in Eq.(1) of the main text to the momentum resolved bandstructure in the normal state of FeSe$_{0.42}$Te$_{0.58}$ measured in angle-resolved photoemission spectroscopy (ARPES) experiments \cite{Tam10}, we obtain the orbital composition of the electronic states on the three Fermi surface sheets [see Fig.~\ref{fig:orb}(a)], as shown in Fig.~\ref{fig:orb}(b).
In particular, we find that states on FS 1 and 2 possess predominant $d_{xz}$- and $d_{yz}$-character, while states on FS 3 possesses large contributions from the $d_{xy}$-orbital and $d_{xz}/d_{yz}$-orbitals]. These results are qualitatively similar to those obtained earlier for LaOFeAs \cite{Gra09}. The angular dependence of the orbital contributions to states along the Fermi surface also explains the anisotropy of the superconducting gap on FS 3. The states on FS 3 that lie along the $\Gamma - M$ direction are predominantly derived from $d_{xy}$-orbitals, while states along the $M-(\pi,\pi)$ direction consist mainly of contributions from either the $d_{xz}$- or $d_{xz}$-orbitals. As the superconducting order parameter of the $d_{xy}$-orbital, $\Delta_{44}=0.38$ meV, is smaller than that of the $d_{xz}$- or $d_{xz}$-orbitals, $\Delta_{11}=\Delta_{22}=0.55$ meV, the superconducting gap on FS 3 along the $\Gamma - M$ direction is smaller than that along the $M-(\pi,\pi)$ direction. This argument is consistent with the angular dependence of the superconducting gap on FS 3 shown in Fig.~1c of the main text, and the results of angle-resolved specific heat measurements \cite{Zeng10}.

\section{Orbital contributions to impurity bound states}

\begin{figure}[!ht]
\renewcommand{\thefigure}{S\arabic{figure}}
\begin{center}
\includegraphics[width=8cm]{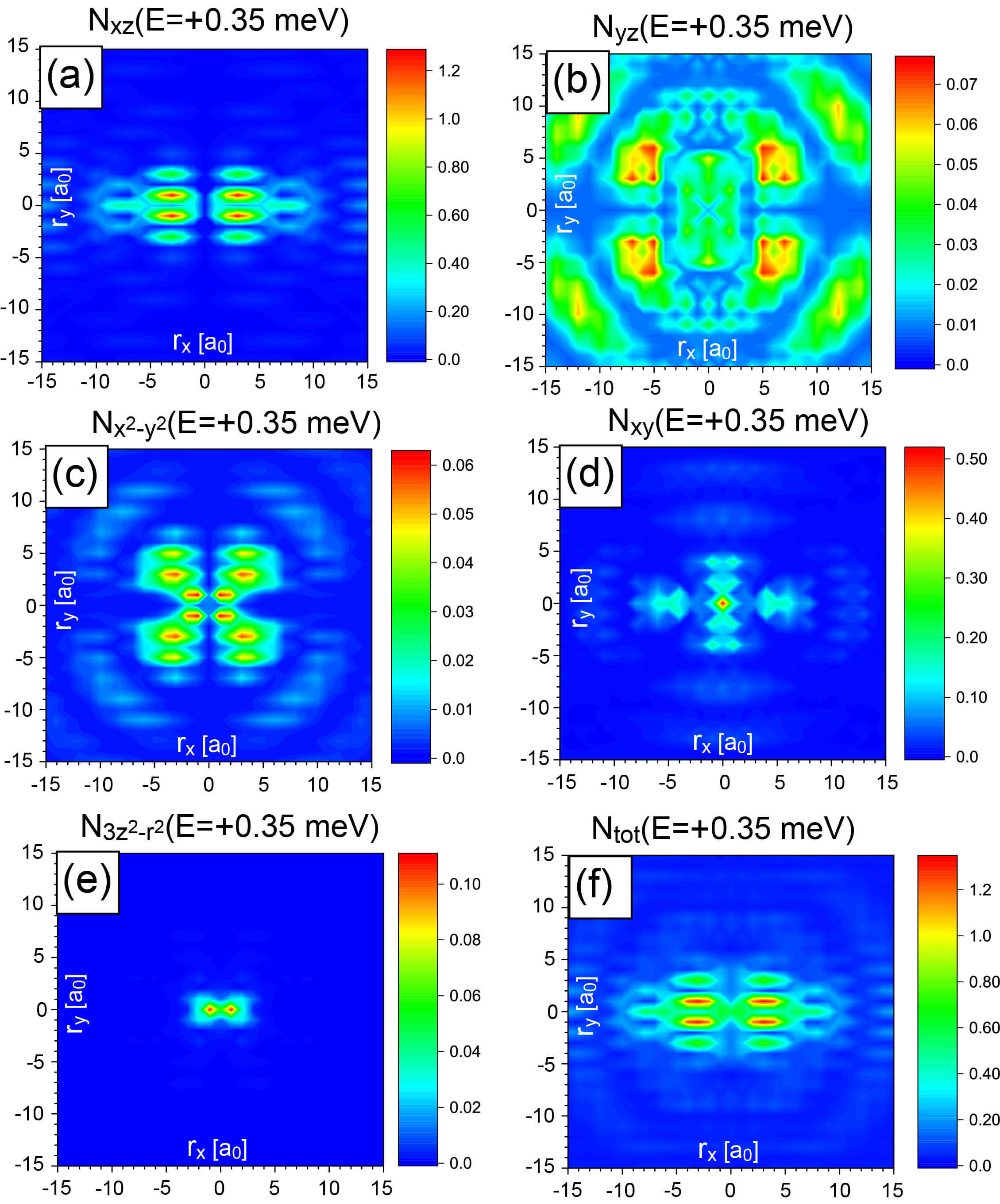}
\caption{Orbitally resolved LDOS at $E=+0.35$ meV for the (a) $d_{xz}$-, (b) $d_{yz}$-, (c) $d_{x^2-y^2}$-, (d) $d_{xy}$-, and (e) $d_{3z^2-r^2}$-orbitals. (f) Total LDOS.
The overall scale of the LDOS (see legends on the right of panels) is the same for all panels.}
\label{fig:orb_defect}
\end{center}
\end{figure}

In Fig.3 of the main text, we showed that a non-magnetic defect, pinning nematic fluctuations, leads to the emergence of an impurity bound state at energies $E_b = \pm 0.35$ meV inside the superconducting gap. In Fig.~\ref{fig:orb_defect} we present the orbitally and spatially resolved LDOS at the energy of the impurity state's hole-branch, $E=+0.35$ meV.
A comparison of the scales of the LDOS for the different orbitals (see legends to the right of panels) reveals that the main contribution to the total LDOS, $N_{tot}$, arises from the $d_{xz}$-orbital. It is therefore likely that the largest contribution to the experimentally measured $dI/dV$ arises from tunneling into the  $d_{xz}$-orbitals (for our choice of $x$- and $y$-axes) and the directionality of the nematic fluctuations.

\section{Evolution of $g({\bf q},E)$  with energy}

\begin{figure}[!ht]
\renewcommand{\thefigure}{S\arabic{figure}}
\begin{center}
\includegraphics[width=8cm]{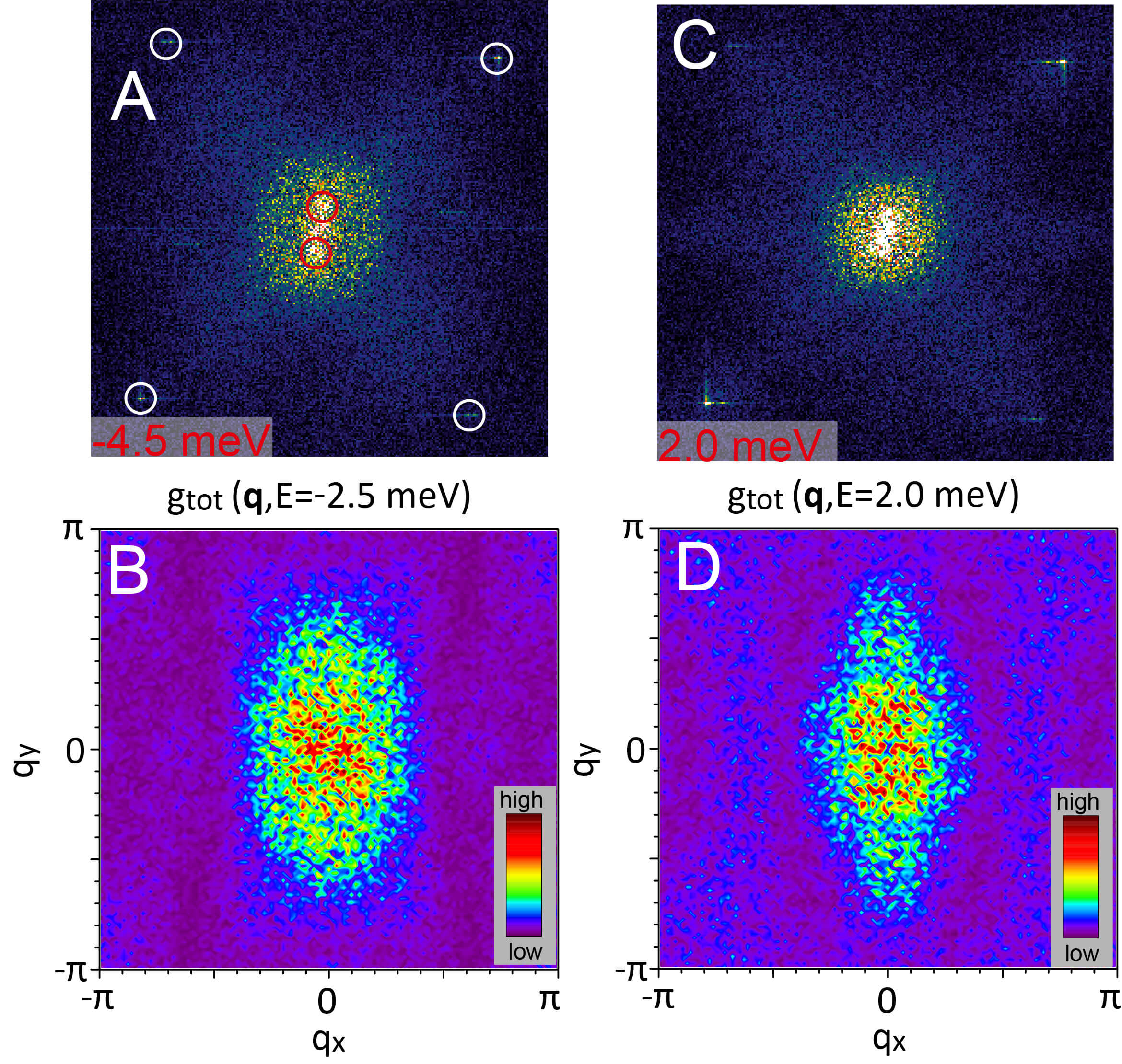}
\caption{Experimentally measured $g({\bf q},E)$ at {\bf A}, $E=-4.5$ meV, and {\bf C}, $E=2.0$ meV in FeSe$_{0.4}$Te$_{0.6}$ from Singh {\it et al.} \cite{Sin15}. Theoretical $g_{tot}({\bf q},E)$ at {\bf B}, $E=-2.5$ meV, and {\bf D}, $E=2.0$ meV. Both sets of spectra show an increasing elongation along the $(0,0) \rightarrow (0,\pi)$ direction for $E \leq 2$ meV.
In {\bf A, C} the white circle marks the position of the Se-Bragg peak. Using a coordinate system where the x- and y-axis are aligned along nearest neighbor Fe-directions, the Se-Bragg peaks are located at ($ \pm \pi/a_{\rm Fe}, \pm \pi/a_{\rm Fe}$). The Se-Bragg peaks are thus rotated by 45 degrees with respect to the Fe-lattice. The elongation of $g({\bf q},E)$ along one of the Fe-Fe-directions is marked with red circles}
\label{fig:qpi_evolve}
\end{center}
\end{figure}

Singh {\it et al.} \cite{Sin15} reported that the Fourier transformed differential conductance, $g({\bf q},E)$, in the superconducting state of FeSe$_{0.45}$Te$_{0.55}$ becomes more elongated along the $(0,0) \rightarrow (0,\pi)$ direction for $E \leq 2$ meV (the latter corresponds to the energy of the coherence peak), as shown in Figs.~\ref{fig:qpi_evolve}A and C. A similar evolution also occurs in the theoretical $g({\bf q},E)$, as demonstrated in Figs.~\ref{fig:qpi_evolve}B and D.
